\documentclass[aps,prl,preprint,tightenlines,superscriptaddress,byrevtex]{revtex4}
\usepackage{graphicx} 
\usepackage{dcolumn}  

\graphicspath{{ps}}

\begin{document}


\preprint{\vbox{ \hbox{   }
                 \hbox{BELLE-CONF-0849}
}}

\title{ \quad\\[0.5cm] Study of $X(3872)$ in $B$ meson decays }


\affiliation{Budker Institute of Nuclear Physics, Novosibirsk}
\affiliation{Chiba University, Chiba}
\affiliation{University of Cincinnati, Cincinnati, Ohio 45221}
\affiliation{Department of Physics, Fu Jen Catholic University, Taipei}
\affiliation{Justus-Liebig-Universit\"at Gie\ss{}en, Gie\ss{}en}
\affiliation{The Graduate University for Advanced Studies, Hayama}
\affiliation{Gyeongsang National University, Chinju}
\affiliation{Hanyang University, Seoul}
\affiliation{University of Hawaii, Honolulu, Hawaii 96822}
\affiliation{High Energy Accelerator Research Organization (KEK), Tsukuba}
\affiliation{Hiroshima Institute of Technology, Hiroshima}
\affiliation{University of Illinois at Urbana-Champaign, Urbana, Illinois 61801}
\affiliation{Institute of High Energy Physics, Chinese Academy of Sciences, Beijing}
\affiliation{Institute of High Energy Physics, Vienna}
\affiliation{Institute of High Energy Physics, Protvino}
\affiliation{Institute for Theoretical and Experimental Physics, Moscow}
\affiliation{J. Stefan Institute, Ljubljana}
\affiliation{Kanagawa University, Yokohama}
\affiliation{Korea University, Seoul}
\affiliation{Kyoto University, Kyoto}
\affiliation{Kyungpook National University, Taegu}
\affiliation{\'Ecole Polytechnique F\'ed\'erale de Lausanne (EPFL), Lausanne}
\affiliation{Faculty of Mathematics and Physics, University of Ljubljana, Ljubljana}
\affiliation{University of Maribor, Maribor}
\affiliation{University of Melbourne, School of Physics, Victoria 3010}
\affiliation{Nagoya University, Nagoya}
\affiliation{Nara Women's University, Nara}
\affiliation{National Central University, Chung-li}
\affiliation{National United University, Miao Li}
\affiliation{Department of Physics, National Taiwan University, Taipei}
\affiliation{H. Niewodniczanski Institute of Nuclear Physics, Krakow}
\affiliation{Nippon Dental University, Niigata}
\affiliation{Niigata University, Niigata}
\affiliation{University of Nova Gorica, Nova Gorica}
\affiliation{Osaka City University, Osaka}
\affiliation{Osaka University, Osaka}
\affiliation{Panjab University, Chandigarh}
\affiliation{Peking University, Beijing}
\affiliation{Princeton University, Princeton, New Jersey 08544}
\affiliation{RIKEN BNL Research Center, Upton, New York 11973}
\affiliation{Saga University, Saga}
\affiliation{University of Science and Technology of China, Hefei}
\affiliation{Seoul National University, Seoul}
\affiliation{Shinshu University, Nagano}
\affiliation{Sungkyunkwan University, Suwon}
\affiliation{University of Sydney, Sydney, New South Wales}
\affiliation{Tata Institute of Fundamental Research, Mumbai}
\affiliation{Toho University, Funabashi}
\affiliation{Tohoku Gakuin University, Tagajo}
\affiliation{Tohoku University, Sendai}
\affiliation{Department of Physics, University of Tokyo, Tokyo}
\affiliation{Tokyo Institute of Technology, Tokyo}
\affiliation{Tokyo Metropolitan University, Tokyo}
\affiliation{Tokyo University of Agriculture and Technology, Tokyo}
\affiliation{Toyama National College of Maritime Technology, Toyama}
\affiliation{Virginia Polytechnic Institute and State University, Blacksburg, Virginia 24061}
\affiliation{Yonsei University, Seoul}
  \author{I.~Adachi}\affiliation{High Energy Accelerator Research Organization (KEK), Tsukuba} 
  \author{H.~Aihara}\affiliation{Department of Physics, University of Tokyo, Tokyo} 
  \author{D.~Anipko}\affiliation{Budker Institute of Nuclear Physics, Novosibirsk} 
  \author{K.~Arinstein}\affiliation{Budker Institute of Nuclear Physics, Novosibirsk} 
  \author{T.~Aso}\affiliation{Toyama National College of Maritime Technology, Toyama} 
  \author{V.~Aulchenko}\affiliation{Budker Institute of Nuclear Physics, Novosibirsk} 
  \author{T.~Aushev}\affiliation{\'Ecole Polytechnique F\'ed\'erale de Lausanne (EPFL), Lausanne}\affiliation{Institute for Theoretical and Experimental Physics, Moscow} 
  \author{T.~Aziz}\affiliation{Tata Institute of Fundamental Research, Mumbai} 
  \author{S.~Bahinipati}\affiliation{University of Cincinnati, Cincinnati, Ohio 45221} 
  \author{A.~M.~Bakich}\affiliation{University of Sydney, Sydney, New South Wales} 
% ...
\collaboration{The Belle Collaboration}
\noaffiliation

\begin{abstract}
We present results on the $X(3872)$, produced in 
$B^+ \to X(3872) K^+$ and $B^0 \to X(3872) K^0_S$ decays where 
$X(3872) \to J/\psi \pi^+ \pi^-$.
We report the first statistically significant observation of 
$B^0 \to X(3872) K^0_S$ and measure the ratio of branching fractions 
to be $\frac{{\cal B}(B^0 \rightarrow X(3872)K^0)}
{{\cal B}(B^+ \rightarrow X(3872)K^+)} = 0.82 \pm 0.22 \pm 0.05$,
consistent with unity. 
The mass difference between the $X(3872)$ states produced in 
$B^+$ and $B^0$ decay is found to be 
$\delta M \equiv M_{XK^+} - M_{XK^0} = (+0.18 \pm 0.89 \pm 0.26)$ 
MeV/$c^2$, consistent with zero. 
In addition, we search for the $X(3872)$ in the decay 
$B^0 \to X(3872)K^+\pi^-$, $X(3872) \to J/\psi \pi^+ \pi^-$. 
We measure ${\cal B}(B^0 \to X(3872) (K^+ \pi^-)_{NR}) 
\times {\cal B}(X(3872) \to J/\psi \pi^+ \pi^-) = 
(8.1 \pm 2.0 ^{+1.1}_{-1.4} )\times 10^{-6}$
and we set the 90\% C.L. limit,
${\cal B}(B^0 \to X(3872)K^{*}(892)^0) 
\times {\cal B}(X(3872) \to J/\psi \pi^+ \pi^-) < 3.4 \times 10^{-6}$. 
The analysis is based on a 605~fb$^{-1}$ data sample collected at the 
$\Upsilon(4S)$ with the Belle detector at the KEKB collider.
\end{abstract}

\maketitle


{\renewcommand{\thefootnote}{\fnsymbol{footnote}}}
\setcounter{footnote}{0}
\section{Introduction}
\par The $X(3872)$ was first observed in the charged $B$-meson 
decay $B^+ \to X(3872) K^+$, 
$X(3872) \to J/\psi \pi^+ \pi^-$ by the Belle 
Collaboration~\cite{x_belle}. 
Its existence has been confirmed by the CDF 
and D0 Collaborations~\cite{x_fermilab} through its inclusive
production in proton-antiproton collisions. 
The discovery mode $B^+ \to X(3872)K^+$ has also been confirmed 
by the BaBar Collaboration~\cite{x_babar}.
The $X(3872)$ mass, combining all measurements in this final state, 
is~\cite{pdg2006}
\begin{equation}
m_{X} = (3871.2 \pm 0.5) \rm MeV \nonumber
\end{equation}
which is at the threshold for the production of the charmed meson
pair $\overline{D}{}^0 D^{*0}$. Recent studies from Belle and
CDF that combine angular information, and kinematic properties 
of the $\pi^+ \pi^-$ pair, strongly favor a $J^{PC} = 1^{++}$ or $2^{-+}$ 
assignment~\cite{x_belle2, x_cdf1, x_cdf2}.
The $X(3872)$ does not appear to be a simple quark model $q\bar{q}$ meson
state: different models have been proposed to explain the nature
of the $X(3872)$ including S-wave $D^0D^{*0}$ molecule 
models~\cite{x_molecule, x_braaten} and various diquark-antidiquark 
models~\cite{x_maiani,x_diquark}. 
The $DD^{*}$ molecule proposal is motivated by the proximity of 
the $X(3872)$ to the $\overline{D}{}^0 D^{*0}$ threshold:
$m_{D^0} + m_{D^{*0}} = 3871.81 \pm 0.25$~MeV~\cite{cleoc, pdg2008}. 
\par In the molecular model, the $X(3872)$ is a $J^{P}=1^+$ state.
Some authors have argued~\cite{x_braaten} that this model, 
together with factorization, heavy-quark and isospin 
symmetries, implies that the ratio of $B^0 \to X(3872) K^0$ to
$B^+ \to X(3872) K^+$ decays is smaller than 0.1; this claim
has recently been challenged~\cite{x_braaten2}.
This ratio is expected to be unity for charmonium as
well as for hybrids ($c\bar{c}g$) and glueballs ($gg$).
\par The diquark anti-diquark model of Maiani {\it et al.}~\cite{x_maiani} 
predicts that the observed $X(3872)$ is one 
component of a doublet of states. In this model, the $X(3872)$ produced 
in charged $B$ meson decays will have a mass that is different from its 
counterpart in neutral $B$ meson decays by 
$\delta M = (7 \pm 2)/\cos (2\theta)~{\rm MeV}$, 
where $\theta$ is a mixing angle that is near $\pm 20^{\circ}$.
\par In order to test the predictions of these models, 
we compare branching fraction and 
$X(3872)$ mass measurements in charged and neutral $B$ decays. 
A previous study performed by BaBar~\cite{x_babar2} using 413~fb$^{-1}$
was not conclusive on these points; this analysis uses a larger sample, 
605~fb$^{-1}$ (657 $\times 10^6 B\bar{B}$ pairs), 
collected  with the Belle detector
at the KEKB asymmetric-energy $e^+e^-$ (3.5~GeV on 8~GeV)
collider~\cite{KEKB} operating at the $\Upsilon(4S)$ resonance.
%
\section{The belle detector}
\par The Belle detector is a large-solid-angle magnetic
spectrometer that consists of a silicon vertex detector (SVD),
a 50-layer central drift chamber (CDC), an array of
aerogel threshold Cherenkov counters (ACC),
a barrel-like arrangement of time-of-flight
scintillation counters (TOF), and an electromagnetic calorimeter (ECL)
comprised of CsI(Tl) crystals located inside
a superconducting solenoid coil that provides a 1.5~T
magnetic field.  An iron flux-return located outside 
the coil is instrumented to detect $K_L^0$ mesons and to identify
muons (KLM).  The detector is described in detail elsewhere~\cite{Belle}.
Two different inner detector configurations were used. For the first sample
of 152 $\times 10^6$ $B\overline{B}$ pairs, a 2.0 cm radius beampipe
and a 3-layer silicon vertex detector (SVD-I) were used;
for the remaining $505 \times 10^6$ $B\overline{B}$ pairs,
a 1.5 cm radius beampipe, a 4-layer silicon detector (SVD-II),
and a small-cell inner drift chamber were used~\cite{Natkaniec}.
%

\section{Selection}

\par Charged tracks are required to originate from the interaction 
point. A likelihood ratio ${\cal R}_{K/\pi} = 
{\cal L}_{K}/({\cal L}_{\pi}+{\cal L}_{K})$, 
where ${\cal L}_{\pi}$ (${\cal L}_{K}$) is the likelihood 
value for the pion (kaon) hypothesis, is built using ACC, 
TOF and CDC ($dE/dx$) measurements. For charged kaons, 
we impose ${\cal R}_{K/\pi}>0.6$ that have an 88\% efficiency 
and a 10\% efficiency for pions.
$K^0_S$ candidates are selected within the $\pi^+ \pi^-$ mass range  
[0.4840, 0.5127] GeV/$c^2$.
Requirements on the $K^0_S$ vertex displacement from the interaction
point and on the difference between vertex and $K^0_S$ flight 
directions are applied. This selection is described in detail
elsewhere~\cite{ks}.

\par We reconstruct $J/\psi$ mesons in the $l^+ l^-$ decay channel
($l = e$ or $\mu$) 
and include bremsstrahlung photons that are within 50 mrad
of either the $e^+$ or $e^-$ tracks (denoted as $e^+ e^- (\gamma)$).
The invariant mass of the $J/\psi$ candidates is required to be 
within $-0.150$ GeV/$c^2$ 
$< M_{e^+ e^- (\gamma)} - m_{J/\psi} < + 0.036$ GeV/$c^2$ and 
$-0.060$ GeV/$c^2 < M_{\mu^+ \mu^-} - m_{J/\psi} < + 0.036$ GeV/$c^2$,
where $m_{J/\psi}$ denotes the $J/\psi$ nominal mass~\cite{pdg2008}, 
and $M_{e^+ e^- (\gamma)}$
and $M_{\mu^+ \mu^-}$ are the reconstructed invariant masses from 
$e^+ e^- (\gamma)$ and $\mu^+ \mu^-$, respectively. 
The $J/\psi$ candidate is then combined with a $\pi^+\pi^-$ pair 
for further analysis: both the $X(3872)$ and the $\psi(2S)$, which 
is used for calibration, decay to this final state. More than 
one $J/\psi \pi^+\pi^-$ combination may be possible at this stage. 
An additional cut is applied on the $M_{\pi^+ \pi^-}$ variable:
$M_{\pi^+ \pi^-} > M(\pi^+ \pi^- J/\psi)- (m_{J/\psi} + m_{\rm cut})$. 
For $B \to J/\psi \pi^+ \pi^- K$ modes 
($B \to J/\psi \pi^+ \pi^- K^+ \pi^-$), we apply the above 
requirement with 
$m_{\rm cut} = 0.2\; \rm GeV$ ($m_{\rm cut} = 0.15\; \rm GeV$). 
For the former case, this cut corresponds to 
$M_{\pi^+\pi^-} > 389$~MeV/$c^2$ for the 
$\psi(2S)$ region and $M_{\pi^+\pi^-} > 575$~MeV/$c^2$ for the $X(3872)$ 
region and reduces significantly the combinatorial background ($\sim 46\%$
in the charged mode) for a reasonable loss of efficiency ($\sim 9\%$).
It also has the property of making the background flat in the 
$M (J/\psi \pi^+ \pi^-)$ variable. 
\par To reduce the combinatorial background 
from $e^+ e^- \to q \overline{q}$ continuum events, 
we require $R_2 < 0.4$ where $R_2$ is the ratio of the 
second to zeroth normalized Fox-Wolfram moments~\cite{foxwolf}, 
and $|\cos \theta_B| < 0.8$, where $\theta_B$ is the polar angle
of the $B$ meson momentum in the center-of-mass (CM) system, 
relative to the $e^+$ beam direction. 
\par $B$ candidates are obtained by combining a $K^+$, a $K^0_S$ or 
$K^+ \pi^-$ candidate with the $J/\psi \pi^+ \pi^-$ candidate. 
We select $B$ candidates using two variables: the energy difference
$\Delta E = E_{B} - E_{\rm beam}^{*} = 
\sum_{i} \sqrt{c^2 p_i^2 + c^4 m_i^2} - E_{\rm beam}^{*}$, and the beam
constrained mass $M_{\rm bc}= (1/c^2) \sqrt{E_{\rm beam}^{*2}-c^2 p_B^2} =
(1/c^2) \sqrt{E_{\rm beam}^{*2}-c^2 (\sum_i p_i)^2}$, where the summation is
over all particles from the $B$ candidate ($p_i$ and $m_i$ are their
CM three-momenta and masses respectively) and $p_B$ is the $B$ 
candidate momentum in the CM frame. 
If more than one candidate is obtained at this stage of the analysis,
the candidate with $\Delta E$ closest to zero is selected. 
Only $B$ candidates with $|\Delta E| < 30$ MeV and 
$M_{\rm bc} < 5.27$ GeV/$c^2$ are considered for further analysis. 
\section{Results}
From this point onwards, we correct the mass measurement using 
the known $J/\psi$ mass, redefining $M (J/\psi \pi^+ \pi^-)$
as $M(J/\psi \pi^+ \pi^-)-M(J/\psi) + m_{J/\psi}$. 
The selection cuts isolate a very pure sample of $B \to \psi(2S) K$,
$\psi(2S) \to J/\psi \pi^+ \pi^-$ decays. These events
are used to calibrate the $M(J/\psi \pi^+ \pi^-)$
resolution and estimate the systematic uncertainty for the $X(3872)$
mass difference. 
\begin{figure}[htbp]
\begin{center}
\begin{tabular}{c}
\includegraphics[width=0.92\textwidth]{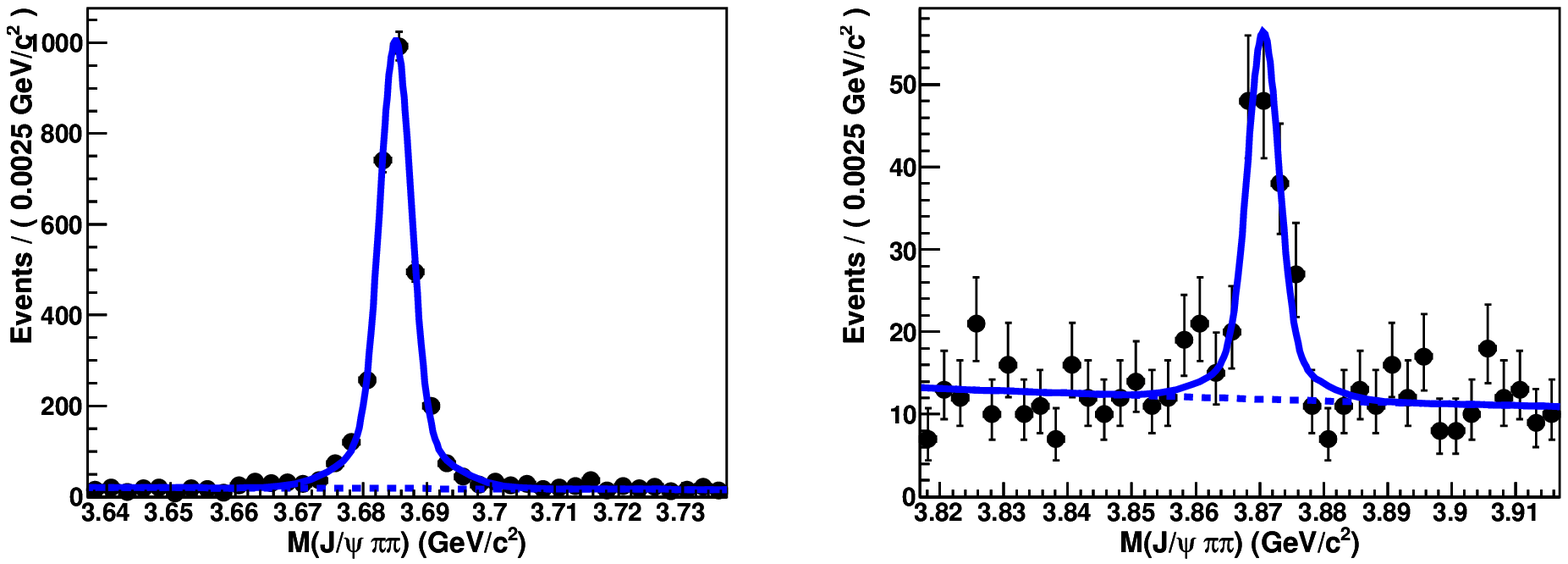} \\
\includegraphics[width=0.92\textwidth]{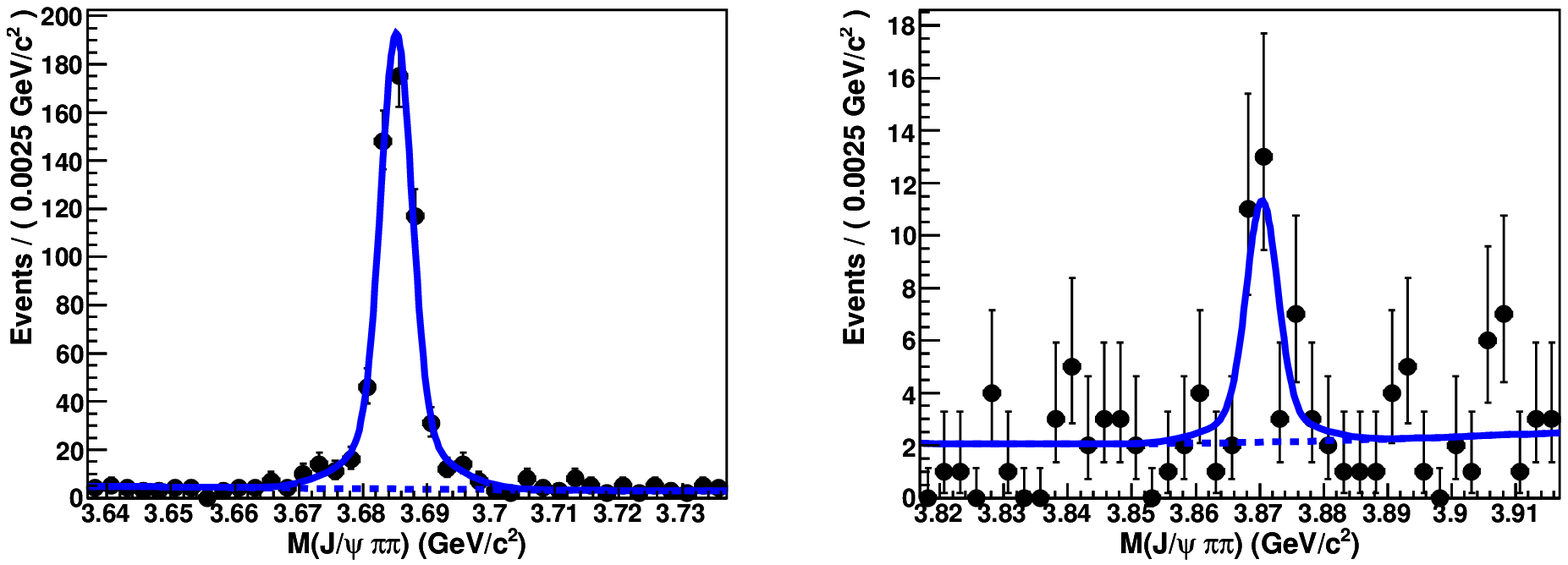} 
\end{tabular}
\end{center}
\caption{The $M(J/\psi \pi^+ \pi^-)$ distribution for the 
$\psi(2S)$ (left) and $X(3872)$ (right) region for charged (top)
and neutral (bottom) $B$ decays. The curve is the result of the
fit described in the text.}
\label{figs_dm_jpsipipik}
\end{figure}
Figure~\ref{figs_dm_jpsipipik} shows the $M(J/\psi \pi^+ \pi^-)$ 
distributions for data near 3686 MeV and 3872 MeV for the charged 
and neutral $B$ modes. 
We perform a fit to the $M(J/\psi \pi^+ \pi^-)$ distribution to
determine the $\psi(2S)$ and $X(3872)$ yields, and the signal 
shape. This fit is performed in the region 
[$m_{J/\psi}+0.54$, $m_{J/\psi}+0.82$] GeV/$c^2$. 
We use the same probability
density function (PDF) for each signal: a sum of two
Gaussians with a common mean. We first perform the fit for
the charged mode, with the $\psi(2S)$ and $X(3872)$ masses and
the two Gaussian widths as free parameters. The width
parameters (mostly determined by the large $\psi(2S)$ peak)
are then fixed, and we perform the fit for the neutral
mode, with only the masses as free parameters. This
procedure allows a clean comparison of the masses in
charged and neutral $B$ decay, for both the $X(3872)$ and the
$\psi(2S)$ control sample.
The signal yields are 
$131.7 \pm 15.0$ and $27.6 \pm 6.6$ for the $X(3872) K^+$ and the 
$X(3872) K^0_S$ modes respectively. The significance is determined
from $-2 \ln ({{\cal L}_0}/{{\cal L}_{\rm max}})$ where 
${\cal L}_0$ and ${\cal L}_{\rm max}$ denote the likelihoods returned
by the fits with the signal yield fixed at zero and at the fitted
value, respectively. This quantity should be distributed as 
$\chi^2 (n_{dof} = 2)$, as two parameters are free for the signal.
The calculated significance is then 12.8$\sigma$ and 5.9$\sigma$, 
respectively.
\begin{table}[ht]
\caption{$X(3872)$ results obtained in the fit described in the text.}
\begin{center}
\begin{tabular}{lcccc}
\hline
\hline
Mode & $\;$ Yield $\;$ & $\;$ $\epsilon (\%)$ $\;$&$\;$ 
Significance ($\sigma$) &
$\;$ ${\cal B} \times {\cal B}(X(3872) \to J/\psi \pi^+ \pi^-)$ \\
\hline
$B^+ \to X(3872) K^+$ & $131.7 \pm 15.0$ & 20.9 & 12.8 & 
$(8.10 \pm 0.92 \pm 0.66) \times 10^{-6}$ \\
$B^0 \to X(3872) K^0$ & $27.6 \pm 6.6$ & 15.2 & 5.9 & 
$(6.65 \pm 1.63 \pm 0.55) \times 10^{-6}$ \\
\hline
\hline
\end{tabular}
\end{center}
\label{tab_results}
\end{table}
\par Using a Monte Carlo (MC) determined acceptance ($\epsilon$), 
the results are summarized in Table~\ref{tab_results} and 
the ratio of branching fractions can then be calculated:
\begin{eqnarray}
\frac{{\cal B}(B^0 \rightarrow X(3872) K^0)}
{{\cal B}(B^+ \rightarrow X(3872) K^+)} & = & 0.82 \pm 0.22 \pm 0.05,
\nonumber
\end{eqnarray}
where we assume the $B^0 \to X(3872) K^0$ transition rate to be equal 
to twice the $B^0 \to X(3872) K^0_S$ rate. 
In this ratio, most of the systematic uncertainties cancel. Therefore 
only the uncertainties due to the $\Upsilon(4S)$ branching 
fractions~\cite{pdg2008} (2.4\%), Monte Carlo statistics 
and MC/data differences are included (Table~\ref{tab_br_syst}). 
The latter source dominates: the differences are due to 
kaon identification (2.2\%) and $K^0_S$ reconstruction 
efficiency (4.5\%).
\begin{table}[ht]
\caption{Summary of the systematic errors in \%.}
\begin{center}
\begin{tabular}{lccc}
\hline
\hline
Source & $X(3872) K^+$ & $X(3872) K^0_S$ & Ratio \\
\hline
$N_{B\overline{B}}$  & 1.4 & 1.4 &  -  \\
Secondary ${\cal B}$ & 1.4 & 1.4 & 2.4 \\
MC statistics        & 0.2 & 0.2 & 0.2 \\
MC decay model       & 2.0 & 2.0 &  -  \\
Kaon ID              & 2.2 &  -  & 2.2 \\
Lepton ID            & 4.2 & 4.2 &  -  \\
Tracking             & 6.0 & 4.8 & 1.2 \\
$K^0_S$ reconstruction &  -  & 4.5 & 4.5 \\
\hline
Total (quadrature) & 8.1  & 8.3 & 5.7 \\
\hline
\hline
\end{tabular}
\end{center}
\label{tab_br_syst}
\end{table}
\par As a check, the ratio was estimated for modes with $\psi(2S)$. 
The signal yields are $2916 \pm 61$ events for $B^+ \to \psi(2S)K^+$ 
and $559 \pm 25$ events for $B^0 \to \psi(2S)K^0_S$ which gives
$\frac{{\cal B}(B^0 \to \psi(2S)K^0 )}{{\cal B}(B^+ \to \psi(2S)K^+)} 
= 0.72 \pm 0.04 \;(\rm stat)$; the systematic error is the same as
that for the $X(3872)$ case, $\pm 0.05$. 
The ratio was also estimated using $\psi(2S) \to l^+ l^-$
decays in the same dataset, finding 
$\frac{{\cal B}(B^0 \to \psi(2S)K^0 )}{{\cal B}(B^+ \to \psi(2S)K^+)} 
= 0.88 \pm 0.05 \;(\rm stat)$.
These two results are in reasonable agreement with the ratio 
calculated from the PDG branching fractions~\cite{pdg2008}, 
$0.96 \pm 0.11$. 
\par The difference between the masses in the charged and 
neutral $B$ modes for $X(3872)K$ is found to be 
$\delta M \equiv M_{XK^+} - M_{XK^0} = (+0.18 \pm 0.89)$~MeV/$c^2$ in data. 
The same calculation is performed for the $\psi(2S)$:
from measured masses of $(3685.12 \pm 0.06)$~MeV/$c^2$ (charged
mode) and $(3685.23 \pm 0.14)$~MeV/$c^2$ (neutral mode), we
find $\delta M = -(0.11 \pm 0.15)$~MeV/$c^2$. There is thus no
significant evidence of $\delta M$ bias; we assign a
conservative systematic error of $\pm 0.26$~MeV/$c^2$ by adding
the central value and one-sigma error for the $\psi(2S)$, and
taking the result as a symmetric error.
The mass difference between the $X(3872)$ states produced in $B^+$ and 
$B^0$ decay is then 
$$
\delta M = (+0.18 \pm 0.89 \pm 0.26)\; {\rm MeV}/c^2
$$
which is consistent with zero.
\par Combining the charged and neutral $B$ samples, we perform a fit 
to the $J/\psi \pi^+ \pi^-$ invariant mass. We correct the 
fitted $X(3872)$ mass by the difference between the $\psi(2S)$ 
world average~\cite{pdg2008} and the mass we measure.
The corrected mass is:
$$
M(X(3872)) = (3871.46 \pm 0.37 \pm 0.07)\; {\rm MeV}/c^2
$$
where the first error is statistical and the second systematic (from 
the $m_{\psi(2S)}$ fit and the nominal mass $m_{\psi(2S)}$).
\par
\begin{figure}[htbp]
\begin{center}
\begin{tabular}{c}
\includegraphics[width=0.92\textwidth]{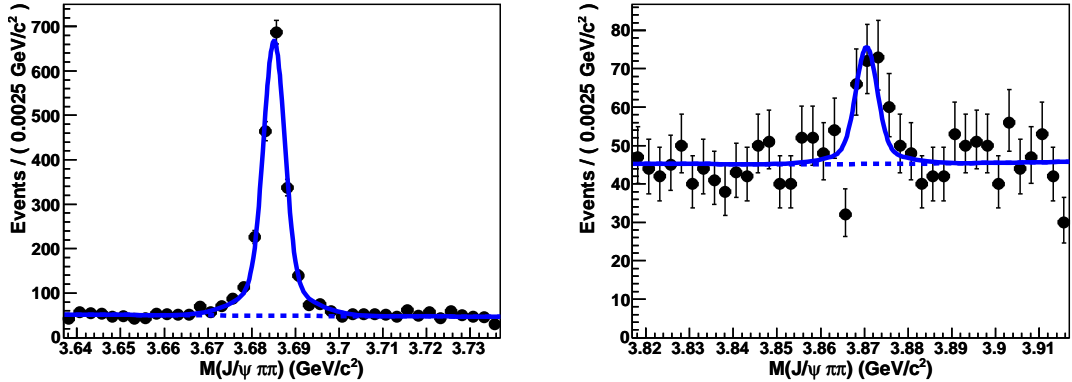} \\
\end{tabular}
\end{center}
\caption{The $M(J/\psi \pi^+ \pi^-)$ distribution for the 
$\psi(2S)$ (left) and $X(3872)$ (right) region for 
$B^0 \to J/\psi \pi^+ \pi^- K^+ \pi^-$.
The curve is the result of the fit described in the text.}
\label{figs_dm_jpsipipikpi}
\end{figure}

\begin{figure}[htbp]
\begin{center}
\begin{tabular}{cc}
\includegraphics[width=0.5\textwidth]{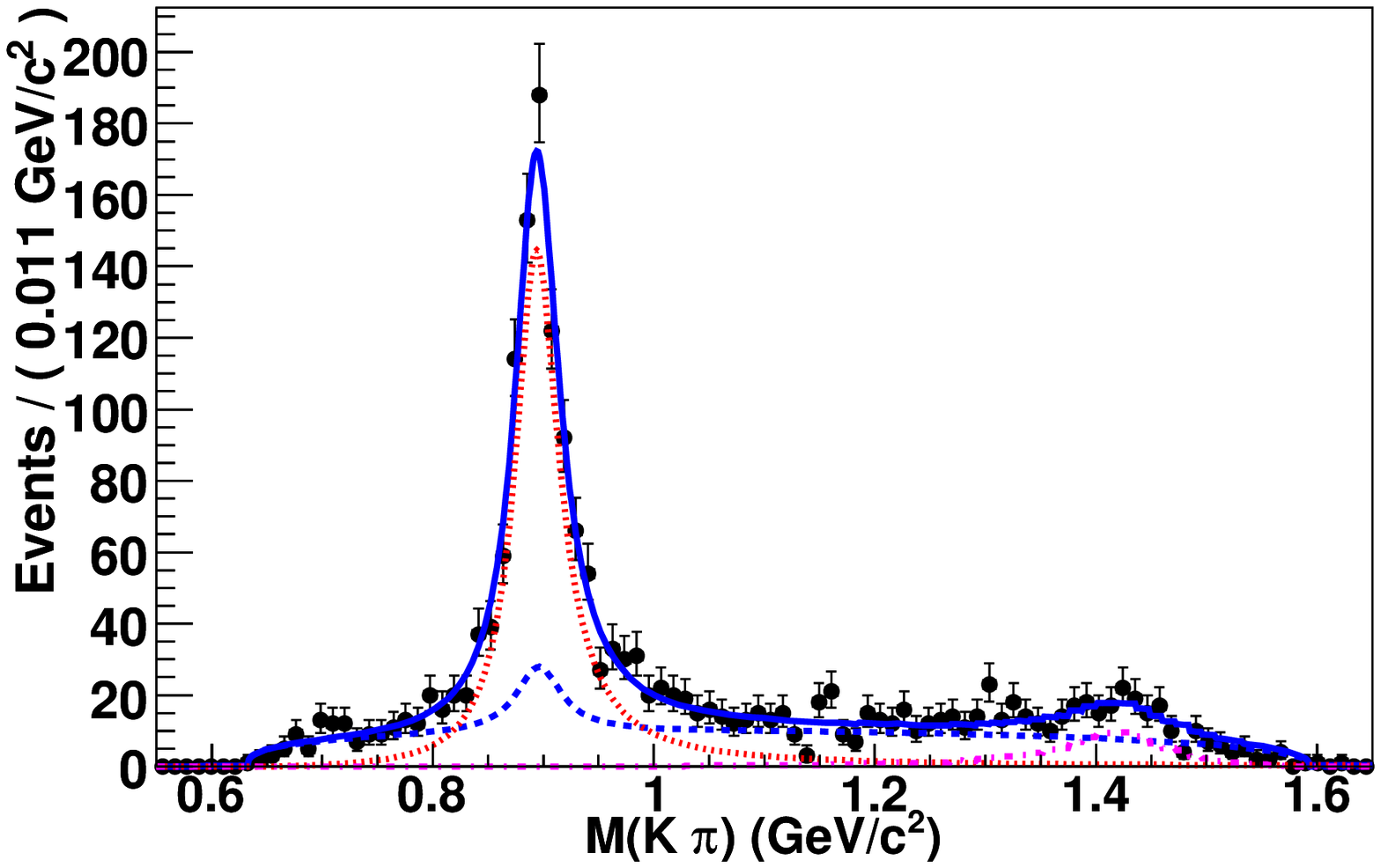} &
\includegraphics[width=0.5\textwidth]{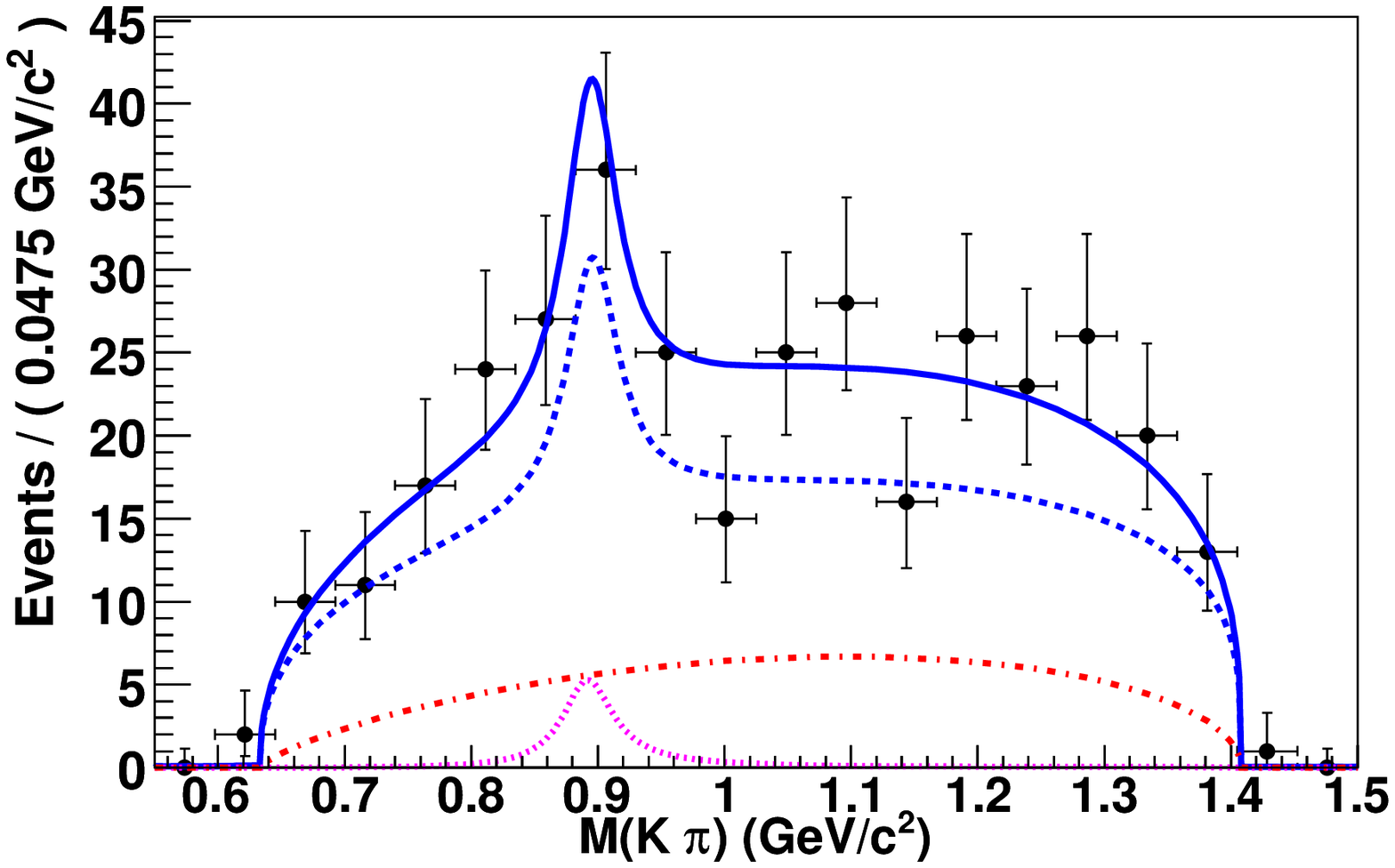} 
\end{tabular}
\end{center}
\caption{The $K\pi$ mass spectrum for the $B \to \psi(2S) K\pi$(left) 
and $B \to X(3872)K\pi$ (right) candidates. 
Left: $B \to \psi(2S) K^{*}(892)^0$ is shown by the dotted red curve, 
$B \to \psi(2S) K^{*}_2(1430)^0$ by the dash-dot magenta curve, 
and the background by the dashed blue curve. Right: 
$B \to X(3872) (K^+ \pi^-)_{NR}$ is shown by the dash-dot red curve, 
$B \to X(3872) K^{*}(892)^0$ by the dotted magenta curve, 
and the background by the dashed blue curve.}
\label{figs_mkpi}
\end{figure}

 A similar fit is performed to $M(J/\psi \pi^+ \pi^-)$ 
for the $B^0 \to (J/\psi \pi^+ \pi^-) K^+ \pi^-$ mode 
(Fig.~\ref{figs_dm_jpsipipikpi}). The signal yield is 
$90 \pm 19$ events.
A fit to the $M_{K \pi}$ distribution is then performed to 
disentangle the $B^0 \to X(3872) K^*(892)^0$ and 3-body 
$B^0 \to X(3872) K^+ \pi^-$ contributions to the final state. 
We select the events within $\pm 7$~MeV around $m_{\psi(2S)}$ 
and $m_{X(3872)}$ and fit their $M_{K \pi}$ distributions 
(Fig.~\ref{figs_mkpi}). 
For the $B^0 \to \psi(2S) K^+ \pi^-$ mode, the PDF for the 
background is the sum of a function 
$(M(K\pi)-m_{K}-m_{\pi})^a \; (m_B - m_{\psi(2S)}-M(K\pi))^b$, 
representing phase space, and a Breit-Wigner function, 
to describe the $K^{*}(892)^0$ contribution, 
obtained from $M(J/\psi \pi^+ \pi^-)$ sidebands
($|M(J/\psi \pi^+ \pi^-)-m_{\psi(2S)} \pm 0.030 | < 0.014$ GeV). 
The PDFs for the signal are a Breit-Wigner PDF with 
a free mean and width to represent the $K^{*}(892)^0$ 
component and a histogram PDF obtained from MC to 
represent the $K^{*}_2(1430)^0$ component.
The signal yield obtained for the $\psi(2S) K^{*}(892)^0$ 
component is $963 \pm 44$ events and corresponds to 
${\cal B}(\psi(2S) K^{*}(892)^0) = 
(5.4 \pm 0.3\;(\rm stat)) \times 10^{-4}$.
This result is in reasonable agreement with the PDG branching 
fraction~\cite{pdg2008}, $(7.2 \pm 0.8) \times 10^{-4}$. 

For the $B^0 \to X(3872) K^+ \pi^-$ mode, the PDF for 
the background is the sum of a phase space function and 
a Breit-Wigner PDF obtained from $M(J/\psi \pi^+ \pi^-)$ 
sidebands
($|M(J/\psi \pi^+ \pi^-)-m_{X(3872)} \pm 0.030 | < 0.014$ GeV).
The background yield is fixed from these sidebands.
The signal is represented 
by two components: a $K^{*}(892)^0$ Breit-Wigner PDF and a 
phase space function obtained from MC. 
The signal yields are $8 \pm 10$ and $81 \pm 20$ events,
respectively. The $K^{*}(892)^0$ contribution 
is not significant and we set the 90\% C.L. limit,
${\cal B}(B^0 \to X(3872)K^{*}(892)^0) 
\times {\cal B}(X(3872) \to J/\psi \pi^+ \pi^-) < 3.4 \times 10^{-6}$. 
A product branching fraction of ${\cal B}(B^0 \to X(3872) 
(K^+ \pi^-)_{NR}) \times {\cal B}(X(3872) \to J/\psi \pi^+ \pi^-) = 
(8.1 \pm 2.0 ^{+1.1}_{-1.4} )\times 10^{-6}$ is also obtained.
For the systematic error contributions, in addition to 
those that enter for $B^+ \to X(3872) K^+$ 
(Table~\ref{tab_br_syst}), we have one more track in the 
final state, limited statistics to fix the background in 
$M_{K\pi}$ ($\pm 10\%$), and possible peaking background 
contributions in $M(J/\psi \pi^+ \pi^-)$,
based on a study of the $\psi(2S) K^+ \pi^-$ calibration mode, 
($^{+0.0}_{-10.6}$\%).
\par 
The result for the $X(3872)$ case is in marked contrast to 
the $\psi(2S)$ case, where the non resonant $B \to \psi(2S) K\pi$ 
component is small and the $B^0 \to \psi(2S) K^{*}(892)^0$ and 
$B^+ \to \psi(2S) K^+$ branching fractions are of comparable 
size. $K^*$ dominance is also found for $B \to J/\psi K \pi$ 
and $\chi_{c1} K \pi$~\cite{pdg2008}.\\

In summary, we report the first statistically significant 
observation of $B^0 \to X(3872) K^0_S$ decay and measure the 
ratio of branching 
fractions to be $\frac{{\cal B}(B^0 \rightarrow X(3872)K^0)}
{{\cal B}(B^+ \rightarrow X(3872)K^+)} = 0.82 \pm 0.22 \pm 0.05$,
consistent with unity. 
The mass difference between the $X$ states
produced in these two decay modes is found to be 
$\delta M \equiv M_{XK^+} - M_{XK^0} = (+0.18 \pm 0.89 \pm 0.26)$ 
MeV/$c^2$, consistent with zero.
In addition, we search for the $X(3872)$ in the decay 
$B^0 \to X(3872)K^+\pi^-$, $X(3872) \to J/\psi \pi^+ \pi^-$. 
We measure ${\cal B}(B^0 \to X(3872) (K^+ \pi^-)_{NR}) 
\times {\cal B}(X(3872) \to J/\psi \pi^+ \pi^-) = 
(8.1 \pm 2.0 ^{+1.1}_{-1.4} )\times 10^{-6}$
and we set the 90\% C.L. limit,
${\cal B}(B^0 \to X(3872)K^{*}(892)^0) 
\times {\cal B}(X(3872) \to J/\psi \pi^+ \pi^-) < 3.4 \times 10^{-6}$. 
%
%
\section*{Acknowledgments}
We thank the KEKB group for the excellent operation of the
accelerator, the KEK cryogenics group for the efficient
operation of the solenoid, and the KEK computer group and
the National Institute of Informatics for valuable computing
and SINET3 network support. We acknowledge support from
the Ministry of Education, Culture, Sports, Science, and
Technology of Japan and the Japan Society for the Promotion
of Science; the Australian Research Council and the
Australian Department of Education, Science and Training;
the National Natural Science Foundation of China under
contract No.~10575109 and 10775142; the Department of
Science and Technology of India; 
the BK21 program of the Ministry of Education of Korea, 
the CHEP SRC program and Basic Research program 
(grant No.~R01-2005-000-10089-0) of the Korea Science and
Engineering Foundation, and the Pure Basic Research Group 
program of the Korea Research Foundation; 
the Polish State Committee for Scientific Research; 
the Ministry of Education and Science of the Russian
Federation and the Russian Federal Agency for Atomic Energy;
the Slovenian Research Agency;  the Swiss
National Science Foundation; the National Science Council
and the Ministry of Education of Taiwan; and the U.S.\
Department of Energy.

\end{document}